# Tachyons and superluminal boosts

**M Ibison**
Institute for Advanced Studies at Austin,
11855 Research Boulevard, Austin TX 78759-2443, USA

E-mail: ibison@earthtech.org

**Abstract.** Some arguments in favour of the existence of tachyons and extensions of the Lorentz Group are presented. On the former, it is observed that with a slight modification to standard electromagnetic theory a single superluminal charge will bind to itself in a self-sustaining circular orbit, suggestive of a (modified) electromagnetic interpretation of the strong force. Symmetries in that theory are used in the subsequent analysis as a starting point in the search for physically-motivated extensions of the Lorentz Group. There is some discussion of the validity of imaginary coordinates in superluminal transformations of frame. The article concludes with some speculation on the implications for faster-than light travel.

## 1. Introduction

From Einstein's principle of relativity that the laws of physics should be the same in all inertial frames one can deduce the existence of a universal constant having the dimensions of a speed relating the measure of time to that of space - though it leaves to empirical observation the determination of the value of this constant (see for example Sexl [1]). 'Special Relativity' is generally taken to mean the principle of relativity *plus* the additional assumption that this universal constant is the speed of light (in vacuum) [2]. It is a commonly asserted alleged inference from - though it is more accurately an addendum to - special relativity that "nothing can go faster than the speed of light". However, neither the principle of relativity nor the constancy of the speed of light deny such a possibility. If believed, Gell-Mann's 'totalitarian principle' - "everything which is not forbidden, is compulsory[1]" - on the contrary demands that superluminal particles (tachyons) must exist. The existence or otherwise of superluminal transformations, however, is a quite separate issue. In the following we discuss both these issues, suggesting reasons to suppose that tachyons exist, and – quite separately - how Special Relativity might be extended to accommodate superluminal transformations.

## 2. Special Relativity

*2.1 Coordinate Transformations*
In modern texts on Special Relativity it is common practice to work out the ramifications of the principle of relativity on the transformation of coordinates between equivalent inertial frames approximately as follows. Denote the time and space coordinates of a pair of events by 4-component vectors $a \equiv (a_0, \mathbf{a})$ and $b \equiv (b_0, \mathbf{b})$, (lower case Latin will denote Euclidean 3-vectors). If a light signal passes from $a$ to $b$, then one can compute its speed:

---

[1] The origin is T. H. White's "The Once and Future King".

$$c^2 = (\mathbf{a}-\mathbf{b})^2/(a_0-b_0)^2. \tag{1}$$

Henceforth all speeds will be normalized with respect to $c$ (and therefore $c \to 1$). It will be useful to write (1) as

$$\tilde{s}gs = 0 \tag{2}$$

where $s \equiv (a_0 - b_0, \mathbf{a}-\mathbf{b})$, the tilde denotes a transpose, and $g$ is the diagonal matrix $g = \text{diag}(1,-1,-1,-1)$. Equivalence between *inertial* frames demands that any coordinate change resulting from relative motion should map straight lines onto straight lines. This dictates that any such transformations of the coordinates must have the form

$$a' = La + d, \quad b' = Lb + d \tag{3}$$

where $L$ is a 4x4 matrix. The constant vector $d$ is just the offset between the origins of the two coordinate systems, and can be set to zero by appropriate alignment. Asserting now the principle of the constancy of the speed of light - independent of inertial frame - (2) and (3) give

$$\tilde{s}'gs' = 0 \Leftrightarrow \tilde{s}\tilde{L}gLs = 0. \tag{4}$$

Since this must be true regardless of the values of $s$, comparing (4) with (2) one obtains constraints on the transformation matrix $L$

$$\tilde{L}gL = \lambda g, \tag{5}$$

where $\lambda$ is a scalar that may depend on the relative velocity. That $\lambda = 1$ is traditionally argued as follows [1,3]: Isotropy of space dictates that $\lambda$ cannot depend on the direction of the velocity and therefore must be an even function: $\lambda = \lambda(\mathbf{v}^2)$. Identify three frames $K$, $K'$ and $K''$ and let $K'$ be at velocity $\mathbf{v}$ relative to $K$, with coordinates transformed by $L(\mathbf{v})$, so that

$$\tilde{L}(\mathbf{v})gL(\mathbf{v}) = \lambda(\mathbf{v}^2)g. \tag{6}$$

Now let $K''$ be at velocity $-\mathbf{v}$ relative to $K'$, so that its coordinates are transformed by $L(-\mathbf{v})$ relative to $K'$, and therefore $L(-\mathbf{v})L(\mathbf{v})$ relative to $K$. Therefore

$$\tilde{L}(-\mathbf{v})\tilde{L}(\mathbf{v})gL(\mathbf{v})L(-\mathbf{v}) = \lambda(\mathbf{v}^2)\lambda(\mathbf{v}^2)g. \tag{7}$$

For the coordinates in $K''$ to be those of $K$ one must have

$$\lambda^2 = 1 \Rightarrow \lambda = \pm 1. \tag{8}$$

Einstein [3] disposes of the possibility $\lambda = -1$ without comment. One thereby arrives at the defining relation for the Lorentz group $L \in M(4,\mathbb{R})$

$$\tilde{L}gL = g. \tag{9}$$

Clearly $L = 1$ (the identity) satisfies (9). Also, for every $L$ satisfying (9), $L^{-1}$ also satisfies (9). Specifically, left-multiplication by $\tilde{L}^{-1}$ and right multiplication by $L^{-1}$ gives

$$\tilde{L}^{-1}gL^{-1} = g. \tag{10}$$

Taking the determinant one has

$$\det(\tilde{L}gL) = \det(g) \Rightarrow \det(\tilde{L})\det(g)\det(L) = \det(g) \Rightarrow (\det(L))^2 = 1 \Rightarrow \det(L) = \pm 1, \quad (11)$$

and this is sufficient to guarantee that an inverse always exists. Matrix multiplication is associative. The real numbers that are the components of the matrices are closed under both addition and multiplication. Consequently the $L$ satisfying (9) form a group under matrix multiplication.

It is notable that though we started only with the requirement (2) that the speed of light be the same in all frames (i.e. (2) is unchanged through homogeneous coordinate transformations) we now have, more generally, that the 'scalar product' $\tilde{a}gb$ (for arbitrary $a,b$) is independent of inertial frame (invariant under Lorentz Transformation). That is,

$$\tilde{a}'gb' = \tilde{a}\tilde{L}gLb = \tilde{a}gb, \quad (12)$$

the last step following from (5) and $\lambda = 1$.

The *proper* Lorentz group of transformations are represented by those $L$ that solve (9) but which are restricted to the subset $\det(L) = +1$ of solutions to (11). These matrices are smoothly connected to the identity and can be written $L = e^\Lambda$, $\Lambda \in M(4,\mathbb{R})$, for which one has $\det L = \exp \operatorname{Tr} \Lambda$ and therefore $\operatorname{Tr} \Lambda = 0$. By contrast, matrices solving (9) but with $\det(L) = -1$ cannot be written as exponentials (of real-valued matrices), and will not, therefore, be smoothly connected to the identity. With this substitution (9) becomes

$$\exp(\tilde{\Lambda})g\exp(\Lambda) = g \Rightarrow \exp(g\tilde{\Lambda}g) = \exp(-\Lambda) \Rightarrow g\tilde{\Lambda}g = -\Lambda \Rightarrow \Lambda g + \widetilde{\Lambda g} = 0. \quad (13)$$

Thus $\Lambda g$ is anti-symmetric and so $\Lambda$ must have the form

$$\Lambda = \begin{pmatrix} 0 & \Lambda_{01} & \Lambda_{02} & \Lambda_{03} \\ \Lambda_{01} & 0 & \Lambda_{12} & \Lambda_{13} \\ \Lambda_{02} & -\Lambda_{12} & 0 & \Lambda_{23} \\ \Lambda_{03} & -\Lambda_{13} & -\Lambda_{23} & 0 \end{pmatrix}. \quad (14)$$

It is deduced that the proper Lorentz group of matrices can be expressed in terms of the generators

$$L = \exp(-\boldsymbol{\omega}.\mathbf{S} - \boldsymbol{\zeta}.\mathbf{K}) \quad (15)$$

where $\boldsymbol{\omega}$ and $\boldsymbol{\zeta}$ are the six free parameters of the transformation; each contains three real scalar degrees of freedom corresponding respectively to rotations and boosts. Each component of $\mathbf{S}$ and $\mathbf{K}$ is a generator for a rotation and a boost respectively:

$$S_1 = \begin{pmatrix} 0 & 0 & 0 & 0 \\ 0 & 0 & 0 & 0 \\ 0 & 0 & 0 & -1 \\ 0 & 0 & 1 & 0 \end{pmatrix}, \quad S_2 = \begin{pmatrix} 0 & 0 & 0 & 0 \\ 0 & 0 & 0 & 1 \\ 0 & 0 & 0 & 0 \\ 0 & -1 & 0 & 0 \end{pmatrix}, \quad S_3 = \begin{pmatrix} 0 & 0 & 0 & 0 \\ 0 & 0 & -1 & 0 \\ 0 & 1 & 0 & 0 \\ 0 & 0 & 0 & 0 \end{pmatrix} \quad (16)$$

and

$$K_1 = \begin{pmatrix} 0 & 1 & 0 & 0 \\ 1 & 0 & 0 & 0 \\ 0 & 0 & 0 & 0 \\ 0 & 0 & 0 & 0 \end{pmatrix}, \quad K_2 = \begin{pmatrix} 0 & 0 & 1 & 0 \\ 0 & 0 & 0 & 0 \\ 1 & 0 & 0 & 0 \\ 0 & 0 & 0 & 0 \end{pmatrix}, \quad K_3 = \begin{pmatrix} 0 & 0 & 0 & 1 \\ 0 & 0 & 0 & 0 \\ 0 & 0 & 0 & 0 \\ 1 & 0 & 0 & 0 \end{pmatrix}. \quad (17)$$

*2.2 Mathematically-motivated extensions of the Lorentz Group*
The above is a brief sketch of the typical deduction of the proper Lorentz Group from the Special Theory of Relativity. Granted *a priori*, the principle of relativity is a powerful tool for the determination of relativistic generalizations of non-relativistic formulae. A good example is the deduction of an expression for the radiated power emitted by a charge in relativistic motion; guided by the 4-vector formalism, one easily deduces the relativistically correct form from the non-relativistic Larmor expression (see for example [4]).

The principle can also reduce the search space of candidate actions when attempting to construct a new physical theory. In particular, one expects the action to be independent of any reference to a preferred inertial frame. More specifically, in so far as the action refers to coordinates (of particles, say), one expects that they appear in the action in such a way that any Lorentz (or Poincaré) transformation leaves the action unchanged. Recalling (12) this implies that physical quantities akin to the time and space coordinates appear only as components of 4-vectors and tensor generalizations thereof.

Many authors have proposed extensions of the Lorentz Group to include superluminal boosts [5-9]. The starting point is (5), but now including the possibility $\lambda = -1$. A purely mathematical approach is to admit the possibility of complex transformations and imaginary coordinates. A superluminal boost in the $x$ direction for example would then be achieved through

$$L = \begin{pmatrix} \frac{1}{\sqrt{v^2-1}} & \frac{v}{\sqrt{v^2-1}} & 0 & 0 \\ \frac{v}{\sqrt{v^2-1}} & \frac{1}{\sqrt{v^2-1}} & 0 & 0 \\ 0 & 0 & i & 0 \\ 0 & 0 & 0 & i \end{pmatrix} \quad (18)$$

where $v$ is an ordinary scalar. Discussion in the literature is concerned with the meaning of the imaginary coordinates and the correct generalization of EM. But here we wish to focus on a different approach: starting from a given action principle that is invariant under (standard) Lorentz Transformations, we find reason to alter the action in order to better accommodate known physics. We then find, 'by accident', support for an extended principle of relativity.

### 3. The bradyon-tachyon Himalayas

We briefly digress to discuss the issue of tachyon rest mass and energy. One may generalize the expression for the inertial energy of a classical point particle to

$$\varepsilon_m = \begin{cases} \dfrac{m_0}{\sqrt{1-v^2}} & bradyons \\ \dfrac{m_0}{\sqrt{v^2-1}} & tachyons \end{cases} \quad (19)$$

so that in both cases the energy still transforms as the time-like component of a 4-vector. That the rest mass of the tachyon is now imaginary is a problem only if it were possible to have a particle pass through light speed. The condition $v = 1$ is referred to by Sudarshan [10] as the 'Himalayas' that keep the two physical realms apart. However, recent work on purely electromagnetic inertial mass [11] suggests instead the generalization

$$\varepsilon_m = \frac{m_0}{\sqrt{|1-v^2|}}, \quad (20)$$

valid for both tachyons and bradyons. Eq. (20) overcomes the objection of imaginary rest masses permitting, conceivably, the deceleration of a tachyon from superluminal to sub-luminal speeds (and a bradyon for sub-luminal to superluminal speeds). There still remains however, the problem that such a particle would have to pass through the speed $v = 1$, apparently requiring infinite energy. This obstacle may not be as insurmountable as it first appears. The Virial Theorem [12] gives that the time-averaged energy of a *closed* system of electromagnetically interacting particles of rest mass $m_j$ is

$$\langle \varepsilon_{\text{total}} \rangle = \sum_j m_j \sqrt{1 - v_j^2} \;, \tag{21}$$

which suggests that light-speed is not, by itself, an insurmountable energetic barrier. In [11] there is given some evidence suggesting that the superluminal generalization of (21) is

$$\langle \varepsilon_{\text{total}} \rangle = \sum_j m_j \, \text{sgn}\left(1 - v_j^2\right) \sqrt{\left|1 - v_j^2\right|} \;. \tag{22}$$

Comparing (20) and (22) one infers the existence of an electromagnetic binding energy

$$\langle \varepsilon_{\text{bind}} \rangle \equiv \langle \varepsilon_{\text{total}} \rangle - \langle \varepsilon_{\text{m}} \rangle = \sum_j \frac{m_j}{\sqrt{\left|1 - v_j^2\right|}} \left(\text{sgn}\left(1 - v_j^2\right)\left|1 - v_j^2\right| - 1\right) = -\sum_j \frac{m_j v_j^2}{\sqrt{\left|1 - v_j^2\right|}} \tag{23}$$

that must be present in a closed system [13]. At $v = 1$, though the total energy is finite (and zero), it appears that the electromagnetic binding must be (negative) infinite, cancelling the Lorentz-boosted inertial energy.

**4. A slightly-modified classical EM**

*4.1 Direct-Action EM*
The action for Maxwell EM can be written

$$I = \int d^4x \left( \frac{1}{8\pi} A \circ \partial^2 A - j \circ A \right) - \sum_i m_i \int dt \sqrt{1 - \mathbf{v}_i^2} \tag{24}$$

where *j* is the total 4 current of all the sources, and *A* the total vector potential (This efficient but rather non-standard action remains gauge invariant due to current conservation – see [14]). Here and throughout the symbol ∘ stands for the Lorentz scalar product

$$a \circ b \equiv a^\mu b_\nu = a^\mu b^\nu g_{\mu\nu} \;. \tag{25}$$

Variation of *A* in (24) gives

$$\partial^2 A = 4\pi j \;. \tag{26}$$

In the direct action version of EM the potential is not an independent degree of freedom and can be eliminated from the action using

$$A(x) = 4\pi \partial^{-2} j = \int d^4x' \delta\left((x - x')^2\right) j(x') \tag{27}$$

where $\partial^{-2}$ denotes convolution with the time-symmetric Green's function. With this the full EM action is now ([15])

$$I = -\frac{1}{2} \int d^4x \int d^4x' \delta\left((x - x')^2\right) j(x) \circ j(x') - \sum_j m_j \int dt \sqrt{1 - \mathbf{v}_j^2} \;. \tag{28}$$

Writing out the current in terms of the sources

$$j(x) = \sum_j e_j \int d\lambda \, \delta^4\left(x_j(\lambda) - x\right) u_j(\lambda) \tag{29}$$

where $u_j(\lambda) \equiv dx_j(\lambda)/d\lambda$, (28) becomes

$$I = -\frac{1}{2}\sum_{j,k} e_j e_k \int d\lambda \int d\lambda' \delta\left(\left(x_j(\lambda) - x'_k(\lambda')\right)^2\right) u_j(\lambda) \circ u'_k(\lambda') - \sum_j m_j \int dt \sqrt{1 - \mathbf{v}_j^2}. \tag{30}$$

Unlike the presentations of Schwarzschild [16], Tetrode [17] and Fokker [18], and Wheeler and Feynman [15,19], self action is present in (30) as a result of the terms $j = k$ in the double-sum. This is in accord with the requirements of the later relativistic development of the direct action theory [20-23], and is discussed in detail in Pegg [24] and by the author [11].

*4.2 Electromagnetic mass*
In both the Maxwell and direct action theories the infinite self-energy resulting from electromagnetic self-action is offset by the mechanical mass $m_j$ so as to give a finite value to their sum, equal to the observed mass. By contrast in [11] is given an account of a version of this theory wherein all mass is entirely electromagnetic, whence the self-action is left completely uncompensated by mechanical mass. The corresponding action is

$$I = -\sum_{j,k} e_j e_k \int d\lambda \int d\lambda' \delta\left(\left(x_j(\lambda) - x'_k(\lambda')\right)^2\right) u_j(\lambda) \circ u'_k(\lambda'). \tag{31}$$

Due to the singular self-action, (31) is regularized by writing

$$I = -\sum_{\sigma=\pm 1}\sum_{j,k} e_j e_k \int d\lambda \int d\lambda' \delta\left(\left(x_j(\lambda) - x'_k(\lambda')\right)^2 - \sigma\Delta^2\right) u_j(\lambda) \circ u'_k(\lambda'). \tag{32}$$

$\Delta$ is a small parameter having units of length introduced to render the self-action initially finite and therefore amenable to analysis. One expands the Euler equations in powers of $\Delta$ and then requires that they are satisfied in the limit that $\Delta \to 0_+$. When $|\mathbf{v}| > 1$ the self-action near $j = k$ $\lambda = \lambda'$ gives rise to a self-force whose leading term has magnitude

$$f_{\text{local}} \sim \frac{e^2 \mathbf{v}^2}{\Delta R \sqrt{\mathbf{v}^2 - 1}} \tag{33}$$

where $R$ is the radius of curvature of the path. There exist additional forces when the trajectory of one particle crosses the light cone of another. These forces will be singular if, at the time of crossing the light cone, a trajectory simultaneously lies on the Cerenkov cone of another, superluminal, particle. In that case it is possible to balance the singular force (33) with an equal and opposite 'distant' force emanating from that crossing point

$$f_{\text{local}} + f_{\text{distant}} = 0, \tag{34}$$

so that the Euler equations are satisfied in the limit that $\Delta \to 0_+$. (It turns out that the Euler equations must be expanded in fractional powers of $\Delta$, though this detail does not impact the discussion here.) The arrangement analyzed in [11] was of two oppositely charged particles in mutually sustaining circular orbits in synchronous superluminal motion about a common centre at relative phase $\pi$. At all times each particle is simultaneously on the light cone and Cerenkov cone of the other, which requirement imposed a constraint on the speed that it satisfy

$$\sqrt{\mathbf{v}^2 - 1} \tan \sqrt{\mathbf{v}^2 - 1} = -1. \tag{35}$$

There are an infinite number of solutions of (35); the first few are given in table 1[2]. For each mode there the system energy and angular momentum are correspondingly quantized.

| mode index | $|\mathbf{v}|$ |
|---|---|
| 1 | 2.972 |
| 2 | 6.202 |
| 3 | 9.371 |
| 4 | 12.526 |
| large $n$ | $n\pi$ |

**Table 1.** Superluminal speeds required of an oppositely charged pair in mutually-sustaining circular motion.

*4.3 A modified direct-action action*

Subsequent to writing [11] it was realized that there is more than one way to apply the 'traditional' (bradyonic) direct-action theory to tachyons. With reference to (31) one observes that whereas the scalar product $u_j(\lambda)\circ u'_k(\lambda')$ is always greater than zero for bradyons, this product may take either sign if either or both of the particles labeled by $j,k$ are tachyons. Consequently there are two different possibilities for generalization to the tachyonic regime that leave the behaviour in the bradyonic regime unchanged. Either (31) is correct as it is, or the physically correct generalization is

$$I = -\sum_{j,k} e_j e_k \int d\lambda \int d\lambda' \delta\left(\left(x_j(\lambda)-x'_k(\lambda')\right)^2\right) \left|u_j(\lambda)\circ u'_k(\lambda')\right|. \tag{36}$$

Recalling that the delta function is an even function of its argument this action is identical to

$$I = -\sum_{j,k} e_j e_k \int d\lambda \int d\lambda' \delta\left(\frac{\left(x_j(\lambda)-x'_k(\lambda')\right)^2}{u_j(\lambda)\circ u'_k(\lambda')}\right) \tag{37}$$

everywhere except possibly at ordinal times $\lambda,\lambda'$ at which

$$u_j(\lambda)\circ u'_k(\lambda') = \pm\infty \quad \left(x_j(\lambda)-x'_k(\lambda')\right)^2 \neq 0 \tag{38}$$

or

$$u_j(\lambda)\circ u'_k(\lambda') = 0 \quad \left(x_j(\lambda)-x'_k(\lambda')\right)^2 = 0. \tag{39}$$

Neither of these conditions can be satisfied if the particles are bradyons, and so (37) is yet another alternative to the action (31), consistent with standard application of the direct action theory.

*4.4 Self-sustaining superluminal motion*

The dynamics resulting from (37) make it an attractive alternative to (31). The balance of forces achieved in [11] was between two particles of opposite sign. For that particular arrangement (circular orbits in synchronous superluminal motion about a common centre at phase difference $\pi$), it was shown in that work that the two particles must be oppositely charged. It can also be shown that for the action (31) a single superluminal particle is incapable of self-sustaining circular motion, i.e. is such is not a solution of the corresponding Euler equations. For a regularized version of the action (37) however,

---

[2] There is an error in [11] involving transformation to a rotating frame. This does not affect (35) however, but only the higher order corrections.

$$I = -\sum_{\sigma=\pm 1}\sum_{j,k} e_j e_k \int d\lambda \int d\lambda' \delta\left(\frac{(x_j(\lambda)-x'_k(\lambda'))^2}{u_j(\lambda)\circ u'_k(\lambda')}+\sigma\Lambda^2\right), \tag{40}$$

one finds that circular motion of a single superluminal particle is now a solution of the corresponding Euler equations in the limit $\Lambda \to 0_+$, provided the velocity satisfies

$$\tan\sqrt{\mathbf{v}^2-1} = \sqrt{\mathbf{v}^2-1}. \tag{41}$$

(The necessity or otherwise of the sum over $\sigma$ is deserving of a proper discussion which is omitted here.) Eq. (41) is the condition that a single charge in circular motion at speed $|\mathbf{v}|$ (always) simultaneously intersects its own light cone and its own Cerenkov cone. The first few solutions are given in table 2. The third column is the number of complete orbits executed between the 'emission' of an EM impulse and reception of the same impulse. The fourth column is the additional fractional part of a cycle between emission and reception expressed as an angle in degrees.

| mode index | $|\mathbf{v}|$ | complete cycles in $\theta$ | residual angle in $\theta$ |
|---|---|---|---|
| 0 | 1.000 | 0 | 0 |
| 1 | 4.603 | 1 | 154.9 |
| 2 | 7.790 | 2 | 165.2 |
| 3 | 10.950 | 3 | 169.5 |
| large $n$ | $(n+\tfrac{1}{2})\pi$ | $n$ | 180.0 |

**Table 2.** Superluminal speeds required of a single charge in self-sustaining circular motion according to the action (40). The mode $n = 1$ is illustrated in figures 1 and 2.

For all but the mode $n = 0$ the Cerenkov forces acting at the present time originate non-locally from past and future times. Acting on the present charge point therefore are always *three* singular forces: the advanced Cerenkov cone influence from a future position on the circular orbit, the retarded Cerenkov cone influence from a past position on the orbit, and the local self-action given by (33) which acts to resist departure from rectilinear motion. Effectively the charge is in three locations at once. The mode $n = 1$ is illustrated in figure 1 as a space-time helix. Points $Q$ and $R$ are respectively future and past points of influence on point $P$. The heavy black lines are the loci of the intersection of the Cerenkov cones from $Q$ and $R$ on the double light cone centered on $P$. The circular orbit of radius $a$ and angular frequency $\omega$ has speed $|\mathbf{v}| = a\omega = 4.603$, which determines the pitch of the helix. The same mode in the x-y plane is depicted in figure 2.

In the case of the mode $n = 0$ the single particle executes self-sustaining circular motion at light speed. In that particular case, and in the limit that $\Lambda \to 0_+$, the forces on the Cerenkov cone coincide with the self-force, acting only locally, i.e. always at the present location of the charge. These behaviors invite the interpretation that $n = 0$ corresponds to a charged lepton, whilst the modes $n > 0$ correspond to the quark structure of baryons, with the implication that each of the three quarks is an 'image' of a single superluminal electron (or positron).

Consider a subluminal test charge in the vicinity of the superluminal orbit. The three 'images' of the superluminal charge at $P$, $Q$, and $R$ will appear to the test charge as spatially separated centers of force. Supposing it is possible to model these as effective charge sources $q_P, q_Q, q_R$, then their sum must still be a unit charge: $q_P + q_Q + q_R = e$. Symmetry gives that $q_Q = q_R$, from which one deduces

$$q_P + 2q_Q = e. \tag{42}$$

Eq. (42) is insufficient to uniquely determine the values of fractional charge of the images, though obviously the charges on the u and d quarks in the proton correspond to the particular solution $q_Q = 2e/3$, $q_P = -e/3$.

## 5. An extended Lorentz Group

*5.1 Scale Invariance*
The above is a rather incomplete sketch of a novel extension of standard EM. Its primary purpose in this document is to provide a physical rather than mathematical motivation to extend the Lorentz group. Returning to the action we first observe that (40) retains the time-reparameterization invariance of (37) only in the limit that $\Lambda \to 0$. This seems to be better than the situation in (32) wherein the regularizing parameter introduces a scale that is absent from (31). Even so, Eq. (31) is invariant to changes in the *magnitude* of the scale, but not the sign; it is not invariant under space-time inversion. This scale magnitude invariance is the well-known 'conformal' invariance of electromagnetism. Traditionally it is broken by the mechanical mass which introduces a definite length scale. It remains intact however in the un-renormalized version of EM. Eq (40) also expresses an additional symmetry: it is invariant under scale changes *and* inversion. As a result, in contrast with the requirement of Special Relativity that the Lorentz norm (2) be invariant under coordinate transformations, it is now sufficient that the ratio of two arbitrary Lorentz scalar products be invariant under coordinate transformations. That is, if *a,b,c,d* are 4-vectors then the action (40) is unchanged by coordinate transformations that leave unchanged the ratio

$$\rho = \frac{\tilde{a}gb}{\tilde{c}gd}. \tag{43}$$

It is easily seen that for homogeneous transformations *L* it is now sufficient that (5) is satisfied

$$\tilde{L}gL = \lambda g \tag{44}$$

where $\lambda$ is an arbitrary real number (of either sign). (Discussion of possible additional transformations that leave (43) unchanged is omitted here.)

Scale invariance seems to be in conflict with experience and common sense. This position is implicit in Einstein's argument in favour of $\lambda = 1$ reproduced above. From this perspective (40), even if it describes real physics, must be incomplete. However it is not clear to the author of this document that the case is closed. Very briefly, scale invariance can be retained if all apparently absolute lengths can be normalized out of the physics. Admittedly this is a tall order. Not only does this require there exist dimensionless expressions relating all particle masses, but it also requires expressions relating particle scales to Cosmological scales, much as the Dirac large number coincidence relates the electron radius to the Hubble radius ([25] is an incomplete sketch of an idea along these lines). Others have argued for restoration of scale (conformal) invariance, see for example [26] for a detailed argument.

*5.2 Superluminal transformations*
If the scale (magnitude) degree of freedom is absorbed into a new matrix $M \equiv L/\sqrt{|\lambda|}$, then (44) can be written

$$\tilde{M}gM = \text{sgn}(\lambda)g. \tag{45}$$

In taking the minus sign in the above, we arrive, albeit by a different route, at the traditional focus of interest for authors whose aim is to extend the Lorentz group to include superluminal transformations [6,8,9,27-30]:

$$\tilde{M}gM = -g. \tag{46}$$

Before proceeding to discuss solutions to (46), we wish to stress that the existence or otherwise of viable superluminal transformations is independent of the existence or otherwise of tachyons. Charges in superluminal motion are already accommodated in standard electromagnetism, including the direct action variant described above. In the Maxwell theory for example there is no problem with asserting a current (29) with speed $|\mathbf{u}(\lambda)/u_0(\lambda)| > 1$. Such a current can be stipulated in the inhomogeneous wave equation (26), to be solved for the potential accordingly. Granted that freedom, it is nonetheless conceivable that nature is such that the equations of physics cannot be formulated so that a change of viewpoint from sub-luminal to super-luminal frame can be accommodated by a homogeneous transformation of the coordinates. The relationship between inertial and accelerated frames is an example that defies such a treatment. With that in mind, let us see what it would take for superluminal transformations to exist.

There are two difficulties to be addressed in the search for meaningful solutions to (46). One is that the matrices $M$ that solve (46) are complex, and some of the coordinates are imaginary. An example is (18). It is the opinion of the author that the appearance of imaginary coordinates signals a failure of the mathematics to properly account for the physics. It can be shown that there exist no real matrices that solve (46) in 1 time and 3 space dimensions. More generally, it can be shown that for real solutions to exist the number of space and time solutions must be equal[3]. In 1+1 D, for example,

$$M = \frac{1}{\sqrt{v^2 - 1}} \begin{pmatrix} 1 & v \\ v & 1 \end{pmatrix} \qquad (47)$$

is a valid solution if $v > 1$. Correspondingly

$$L = \begin{pmatrix} 1 & v \\ v & 1 \end{pmatrix} \qquad (48)$$

is a solution of (44), where the scale factor is $\lambda = v^2 - 1$. This transformation is valid for all $v$ – superluminal and sub-luminal. Given that we observe 3 spatial dimensions therefore, in order that real superluminal transformations exist there must be at least 3 space and 3 time dimensions, a fact that has been recognized and discussed by Cole [5,31,32], and by Maccarrone and Recami [7].

The second difficulty is that neither (44) nor (46) define a group. The identity is clearly not a solution of (46), so this equation does not define a group. Upon taking the determinant of (44) we find

$$\det(\tilde{L}gL) = \det(\lambda g) \Rightarrow -\det(L)^2 = -\lambda^4 \Rightarrow \det(L) = \pm\lambda^2. \qquad (49)$$

The determinant of the matrix $L$ vanishes when $\lambda = 0$ and so has no inverse. In 1+1 D, with reference to (48), this is the state of affairs when $v = 1$, i.e. when the transformation is at light speed. The cause of the mathematical pathology is seen to be that the light speed transformation maps the x, t plane onto a line (or pair of lines) $x = \pm t$. Since there is no unambiguous mapping from a line to a plane there can be no inverse transformation. From the perspective of (44) and in the context the direct-action (40) the normalization procedure leading to (45) is invalid at $\lambda = 0$, so that (45) cannot be considered as fundamental in the sense of (44). By contrast, outside the context of (40), (45) may be considered as a defining relation for an extended Lorentz group, provided the special case $\lambda = 0$ is excluded. This appears to be the position of previous authors who have written on superluminal transformations. In the equation

$$\tilde{M}gM = \pm g \qquad (50)$$

the matrices $M$ form a group because the determinant never vanishes. This conclusion applies in 3+1 D, (where the $M$ are complex in the case of the negative sign), and in 3+3 D in which solutions exist for real $M$ in the case of the negative sign.

---

[3] The author is very grateful to Eric Katerman for this proof. We hope to report these findings in a future publication.

The $L$ satisfying (44) form a group provided the case $\lambda = 0$ is excluded, just as real numbers form a group under multiplication provided zero is excluded. Though this may be a satisfactory construction from a mathematical point of view, it is less so physically. Because the Lie group is non-connected one cannot successively apply infinitesimal generators from the identity (which solves (9)) to arrive at solutions of (46). The lack of a connectedness suggests that achieving superluminal motion is not simply a matter of (smooth) acceleration through light speed. If so, with respect to the sub-luminal domain the superluminal domain may as well be regarded as discrete symmetry in the manner of time or parity reversal. With $\lambda = 0$ excluded, though we can say what a sub-luminal event would look like from a superluminal frame, we cannot countenance an inhabitant of the sub-luminal domain boarding a space craft and witnessing that event from a superluminal frame. From an engineering perspective therefore, the bradyon-tachyon Himalayas still exist.

## 6. Final remarks

The relation (50) suffices if the goal of the extension of the Lorentz Group is to provide a unified physical treatment of the sub and superluminal domains. If the coordinates in both domains must be real, then one additionally requires that there be three time and three space dimensions. If however the goal of the extension of the Lorentz Group is to achieve real transformations valid in the sub-luminal and superluminal domains, *and* at light speed, then a further extension is required. One possibility is to add another space dimension so that the matrix transformation corresponding to light speed boost is no longer singular. In 1+1 D for example, (48) would be taken to represent a 2x2 sub-matrix of a 3x3 matrix where entries in the third row and column of the latter are very small, but not zero. With this, one could then hope to write equations that when subject to homogeneous transformations correctly described the same physics as viewed from any of the three domains - in the sub-luminal and superluminal domains, *and* at light speed. Presumably it would also open the door to acceleration through light speed. The price paid for this extension would that the Lorentz boost would no longer be an exact invariance in 3+1 or 3+3, though in its place would be another more exact invariance involving all 7 coordinates in 3+3+1 D.

## 7. Acknowledgements

The author is grateful to Harold Puthoff and Eric Katerman for useful and entertaining discussions on the issues discussed here. Whatever inaccuracies are present, however, are the sole responsibility of the author.

## 8. References

[1] Sexl R U 2000 *Relativity, Groups, Particles : Special Relativity and Relativistic Symmetry in Field and Particle Physics* (Wien: Springer-Verlag)
[2] Rindler W 1991 *Introduction to Special Relativity* (Oxford, UK: Oxford University Press)
[3] Einstein A 1966 *The Meaning of Relativity* (Princeton , USA: Princeton University Press)
[4] Jackson J D 1998 *Classical Electrodynamics* (New York: John Wiley & Sons, Inc.)
[5] Cole E A B 1977 Superluminal transformations using either complex space-time or real space-time symmetry *Nuovo Cimento A* **40** 171-80
[6] Corben H C 1974 Imaginary Quantities in Superluminal Lorentz Transformations *Lett. Nuovo Cimento* **11** 533-6
[7] Maccarrone G D and Recami E 1984 The introduction of superluminal Lorentz transfoarmtions: a revisitation *Found. Phys.* **14** 367-408
[8] Mignani R and Recami E 1973 Generalized Lorentz Transformations in Four Dimensions and Superluminal Objects *Nuovo Cimento* **14A** 169-89
[9] Recami E 1978 An introductory view about superluminal frames and tachyons *Tachyons, Monopoles, and Related Topics* (Amsterdam: North-Holland) pp 3-25
[10] Bilaniuk O M, Deshpande V K, and Sudarshan E C G 1962 *Am. J. Phys.* **30** 718-23
[11] Ibison M 2006 Un-renormalized Classical Electromagnetism *Annals of Physics* **321** 261-305
[12] Landau L D and Lifshitz E M 1980 *The Classical Theory of Fields* (Oxford, UK: Pergamon Press)

[13] Ibison M 2007 Emergent gravity from direct-action EM in a toy universe of electrons and positrons *submitted to Phys. Lett. A*
[14] Itzykson C and Zuber J-B 1985 *Quantum field theory* (New York: McGraw-Hill)
[15] Wheeler J A and Feynman R P 1949 Classical electrodynamics in terms of direct interparticle action *Rev. Mod. Phys.* **21** 425-33
[16] Schwarzschild K 1903 ? *Gottinger Nachrichten* **128** 132
[17] Tetrode H 1922 Ûber den wirkungszusammenhang der welt. Eine erweiterung der klassischen dynamik *Z. Phys.* **10** 317-28
[18] Fokker A D 1929 Ein invarianter Variationssatz für die Bewegung mehrerer elektrischer Massenteilchen *Z. Phys.* **58** 386-93
[19] Wheeler J A and Feynman R P 1945 Interaction with the absorber as the mechanism of radiation *Rev. Mod. Phys.* **17** 157-81
[20] Davies P C W 1971 Extension of Wheeler-Feynman quantum theory to the relativistic domain I. Scattering processes *J. Phys. A* **4** 836-45
[21] Davies P C W 1972 Extension of Wheeler-Feynman quantum theory to the relativistic domain II. Emission processes *J. Phys. A* **5** 1024-36
[22] Hoyle F and Narlikar J V 1969 Electrodynamics of Direct interparticle Action. I. The Quantum Mechanical Response of the Universe. *Annals of Physics* **54** 207-39
[23] Hoyle F and Narlikar J V 1971 Electrodynamics of Direct interparticle Action II. Relativistic Treatment of Radiative Processes *Annals of Physics* **62** 44-97
[24] Pegg D T 1975 Absorber Theory of Radiation *Rep. Prog. Phys.* **38** 1339-83
[25] Ibison M 2002 A ZPF-Mediated Cosmological Origin of Electron Inertia *Gravitation and Cosmology: From the Hubble Radius to the Planck Scale* eds G Hunter et al (Dordrecht: Kluwer Academic)
[26] Burbidge G, Hoyle F, and Narlikar J V 1999 A Different Approach to Cosmology *Physics Today* April 38-46
[27] Corben H C and Honig E 1975 Behavior of Electromagnetic Charges under Superluminal Lorentz Transformations *Lett. Nuovo Cimento* **13** 586-8
[28] Corben H C 1978 Electromagnetic and Hadronic Properties of Tachyons *Tachyons, Monopoles and Related Topics* ed E Recami (Amsterdam: North-Holland) pp 31-41
[29] Corben H C 1975 Tachyon Matter and Complex Physical Variables *Nuovo Cimento* **29A** 415-26
[30] Recami E and Mignani R 1974 Classical Theory of Tachyons (Special Relativity Extended to Superluminal Frames and Objects) *Rivista del Nuovo Cimento* **4** 209-90
[31] Cole E A B 1980 Center-of-mass frames in six-dimensional special relativity *Lett. Nuovo Cimento* **28** 171-4
[32] Cole E A B 1978 Subluminal and superluminal transformations in six-dimensional special relativity *Nuovo Cimento B* **44** 157-66

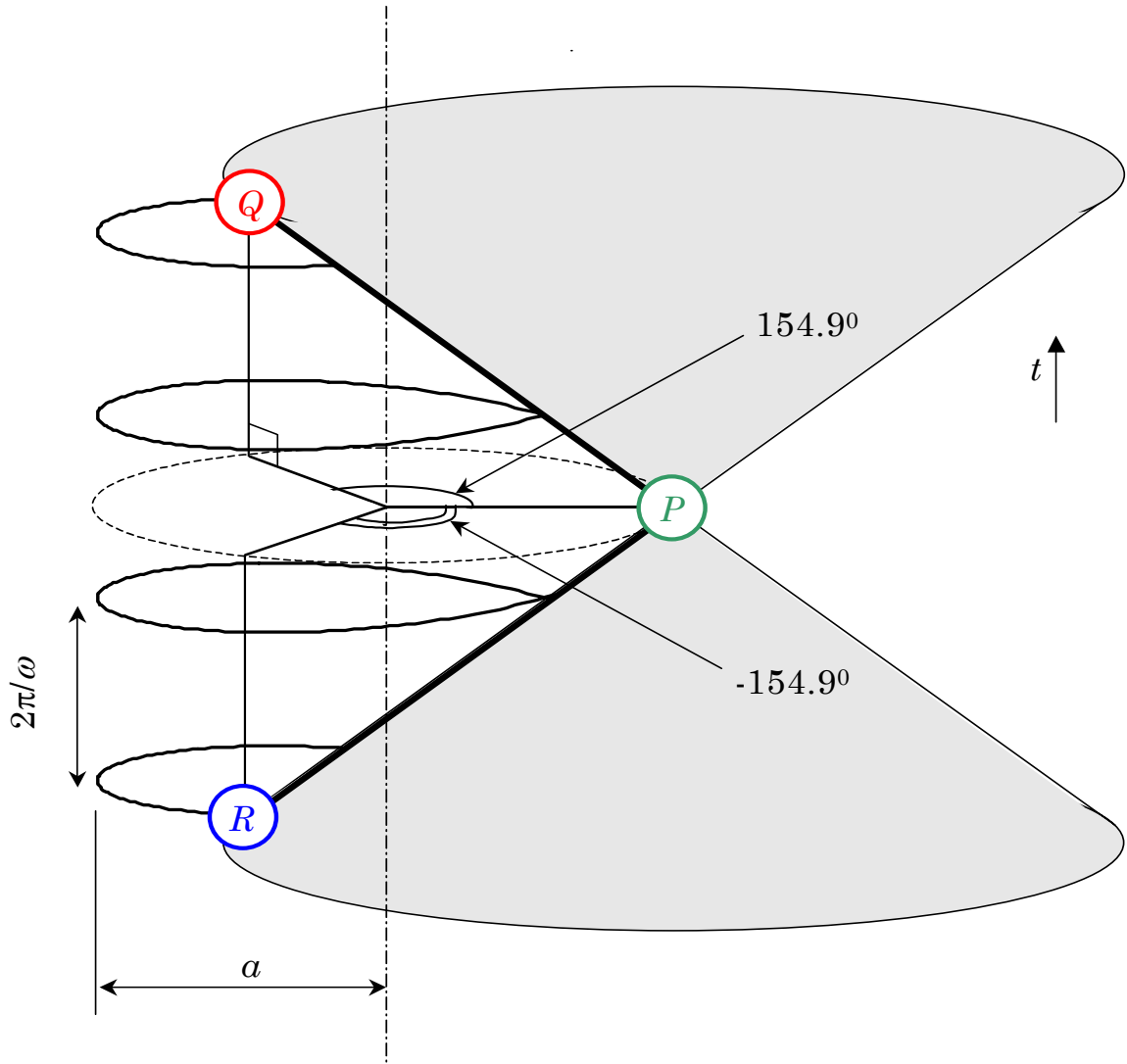

**Figure 1.** Circular motion of a single charge in superluminal motion depicted as a space-time helix. The heavy black lines are the intersections of the Cerenkov cones with the light cones and are the loci of singular electromagnetic forces. The nominal present location of the charge is $P$. The points $Q$ and $R$ are the future and historical location (with respect to $P$) of the same charge, and which act on $P$ via advanced and retarded influences respectively. The allowed (self-sustaining) speeds that satisfy the Euler equations for the action (40) are quantized. Shown here is the mode $n = 1$.

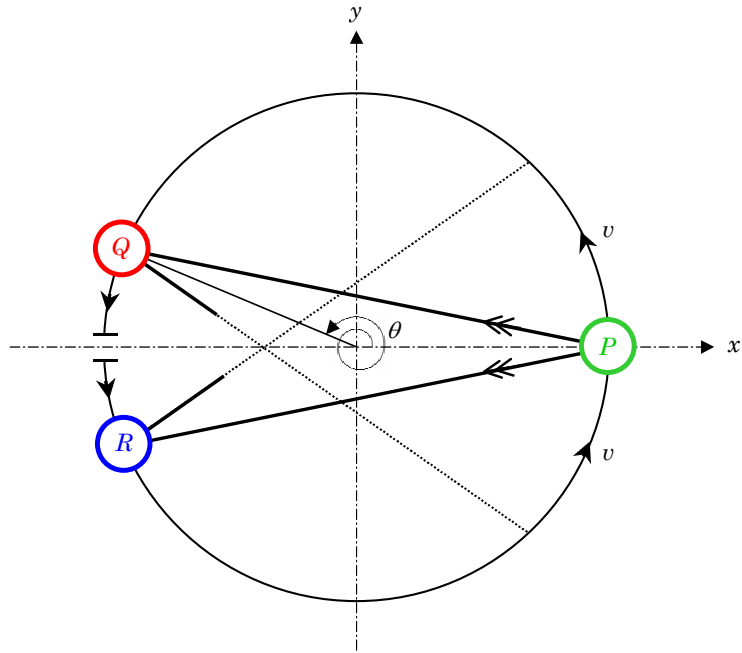

**Figure 2.** The mode $n = 1$ of a singular charge in self-sustaining motion in the x,y plane. $\theta$ is the relative angle between the sources of singular force. Due to the modified action (40), the (same) charge at $Q$ and $R$ appears oppositely charged to that at $P$, giving rise to a net attractive force with a resultant in the negative $x$ direction.